\documentclass{article}[11.8pt]

\usepackage{graphicx}
\usepackage{fullpage,amsmath,amssymb,latexsym}
\usepackage{multirow}
\usepackage{natbib}

\begin{document}
\title{On the Evolution and Survival of Protoplanets \\
Embedded in a Protoplanetary Disk}

\author{A. Vazan and R. Helled\\
Department of Geophysics, Atmospheric, and Planetary Sciences\\
   Tel-Aviv University, Israel}
\date{}
\maketitle 

\begin{abstract}
We model the evolution of a Jupiter-mass protoplanet formed by the disk instability mechanism at various radial distances accounting for the presence of the disk. Using three different disk models, it is found that a newly-formed Jupiter-mass protoplanet at radial distance of $\lesssim$ 5-10 AU cannot undergo a dynamical collapse and evolve further to become a gravitational bound planet. We therefore conclude that {\it giant planets, if formed by the gravitational instability mechanism, must form and remain at large radial distances during the first $\sim$ 10$^5-10^6$ years of their evolution}. The minimum radial distances in which protoplanets of 1 Saturn-mass, 3 and 5 Jupiter-mass protoplanets can evolve using a disk model with $\dot{M}=10^{-6} M_{Sun}/yr$ and $\alpha=10^{-2}$ are found to be 12, 9, and 7 AU, respectively. The effect of gas accretion on the planetary evolution of a Jupiter-mass protoplanet is also investigated.  It is shown that gas accretion can shorten the pre-collapse timescale substantially. Our study suggests that the timescale of the pre-collapse stage does not only depend on the planetary mass, but is greatly affected by the presence of the disk and efficient gas accretion.
\end{abstract}


\section{Introduction}
The formation mechanism for giant planets is still not fully understood. Currently there seems to be a general agreement that the standard model for giant planet formation is {\it core accretion}, where giant planets are formed by accumulation of solid bodies followed by accretion of a gaseous envelope (e.g., Pollack et al., 1996, Hubickyj et al., 2005). However, the {\it gravitational (disk) instability} model in which gaseous planets are formed as a result of gravitational instabilities in the disk surrounding the young star (e.g., Boss, 1997; Mayer et al., 2007), might be required to explain the formation of massive gaseous planets at very large radial distances, the existence of giant planets around low-metallicity stars, giant planets orbiting very young stars, etc. \par 

Giant planet formation by disk instability is still a subject for ongoing research (e.g., Boss, 1997, 2011; Boley, 2009, Boley et al, 2010; Nayakshin, 2011, Rafikov, 2009) and the conditions under which planets can form are still not well determined.  One important aspect of giant planet formation in the disk instability scenario is the survival of the planets. Even if clumps do manage to form it is not clear whether these objects can remain gravitationally bound and evolve further to become gravitationally bound planets (Rafikov, 2007; D'Angelo et al, 2011; Durisen et al., 2007; Cai et al., 2008).  Most of the models concentrate on the actual formation of clumps in disks, and often cannot follow the planetary evolution in detail due to numerical limitations, although, effort in that direction is in progress (e.g., Nayaskin, 2011; Hayfield et al., 2011). In this paper, we follow in detail  the evolution of protoplanets when the presence of the protoplanetary disk is included, and investigate under what conditions such planetary objects manage to survive and evolve to become gas giant planets. We also explore the effect of gas accretion on the planetary evolution.  \par

\section{Evolution of an Embedded Jupiter-Mass Protoplanet}\label{dstnc}
\subsection{The Evolution Model}

The evolution of a  planetary object formed by a gravitational instability in a
protoplanetary disk includes three phases (e.g., DeCampli and Cameron, 1979; Bodenheimer et al., 1980): 
(i) Once a clump is formed it contracts quasi-statically having low 
internal temperatures on the order of a few hundred Kelvin. Hydrogen is in
molecular form (H$_2$), and the radius of the protoplanet is of the order of the Hill radius. (ii) When the central temperature reaches $\sim$ 2000 K, the molecular hydrogen starts to 
dissociate initiating a dynamical collapse of the entire protoplanet, ending
only when the radius has decreased to a few times the present Jupiter radius. 
(iii) The protoplanet contracts quasi-statically and cools on a timescale of $\sim 10^9$ years. \par

The pre-collapse stage is the most crucial for clump survival, due to the extended planetary configuration and the influence of the protoplanetary disk, and we therefore focus on that evolutionary phase. 
To model the pre-collapse evolution we use a static, 1D (i.e. no rotation) planetary evolution code that solves the standard stellar evolution equations (Helled et al. 2006, 2008, Kovetz et al. 2009) with modified boundary conditions to include the presence of the disk. The protoplanet is assumed to consist of only hydrogen and helium with a proto-solar ratio. Since the protoplanet is no longer assumed to be isolated we insert the influence of the disk (at a given radial location) into the evolution calculation. 
The modifications are as follows: (a) irradiation flux (via the disk's temperature due to the opaqueness of the disk) is included in the luminosity boundary condition (b) the disk's pressure is taken into account in the photospheric boundary condition. Therefore, the boundary conditions of the photosphere are \begin{equation} L=4\pi r^2\sigma_{SB}\left( T^4 - g(\tau_s) T_{irr}^4\right), \qquad  \kappa(p-p_0)=\frac{Gm\tau_s}{r^2}
\end{equation} 
where $T_{irr}$ is the disk's temperature, $ \kappa(\rho,T)$ is the opacity, $\tau_s$ is the optical depth taken to be unity, and $g(\tau_s)=\frac{3}{2}(1-\frac{1}{2}e^{-\tau})$ (Kovetz et al.,1988). 
$p_0$ is the disk's pressure, calculated from disk's density assuming an ideal gas, and the remaining symbols are in standard notation. The planetary opacity is computed for a solar-composition using opacity tables that were kindly provided by P. Bodenheimer, based on the work of Pollack et al. (1985). They include both gas and grain opacity, the later being based on the size distribution relevant for interstellar grains. For simplicity, the effect of grain growth and settling on the planetary opacity and the evolution (Helled and Bodenheimer, 2011) is not included.   \par

We first use a disk model kindly provided by Bell based on the work of Bell et al. (1997). This disk model (hereafter {\it disk1}) corresponds to a 1 M$_{Sun}$ star, with viscosity parameter $\alpha=10^{-2}$, and mass accretion rate of $\dot{M}=10^{-6} M_{Sun}/yr$ (Fig 1. of Bell et al., 1997). 
The photospheric boundary conditions of the protoplanet are determined by the disk model, which provides density and temperature as a function of radial distance, and as a result, the pressure can be calculated assuming ideal gas.  Since the protoplanetary disk is optically thick at early stages direct irradiation from the star can be neglected (e.g., D'Alessio 1999) and we therefore account only for the disk's temperature. As a second step we check the sensitivity of the results using two other disk models (see section 2.4). \par

For simplicity, we do not consider planetary migration during our calculation. However, as discussed below, we assume that the protoplanets have not formed at small radial distances where the disk is gravitationally stable but arrived there due to rapid inward migration or an alternative physical process. In addition, we assume that the disk's physical properties are unchanged during the pre-collapse evolution. Since the disk's temperature and pressure are decreasing with time, our results provide {\it upper bounds} for the pre-collapse timescale since as the disk evolves, the boundary conditions on the planetary photosphere are expected to relax. \par

The initial model of the protoplanet at a given radial distance is derived assuming an adiabatic internal structure, 
and photospheric temperature and density that are similar to (or slightly higher than) those of the disk. 
First, we model the planetary evolution of a 1 M$_J$ protoplanet, where M$_J$ is Jupiter's mass, for radial distances between 6 and 50 AU. The physical properties of the disk and the initial model of the protoplanet for radial distances of 20 and 50 AU are presented in table 1. In section 2.3 the evolution of protoplanets of various masses is investigated.  \par 

\subsection{Model Results}
The initial radii used in this work were chosen to be extended, with photospheric temperatures and pressures that are similar to those of the surrounding disk as expected for protoplanets formed by gravitational instabilities (see e.g., Boss, 2002; Helled and Bodenheimer, 2010, 2011, and references therein for details). As a result, using very different initial effective temperatures might be out of context as we discuss below. In addition, the initial planetary configuration has only a small effect on the pre-collapse evolution, and it is found that the 'memory' of the initial model lasts for only a few hundred years. \par 

\begin{figure}[ftb]
\begin{center}
\includegraphics[angle=0, width=13cm]{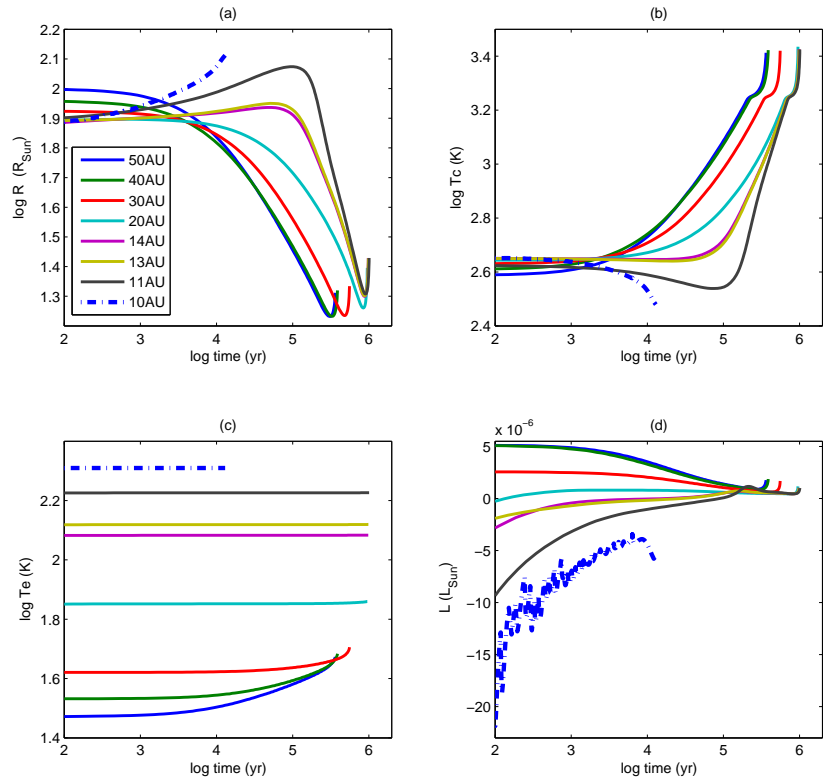}
\caption{Pre-collapse evolution for a Jupiter-mass protoplanet for radial distances between 10 and 50AU. The colors correspond to different radial distances (see legend). Shown are the radius $R$ (a), central temperature $Tc$ (b), effective temperature $Te$ (c), and luminosity $L$ (d). }\label{distfig}
\end{center}
\end{figure}

The change in the physical properties of a newly-formed Jupiter-mass protoplanet as a function of time at the various radial distances is presented in figure \ref{distfig}. The evolution is shown until the protoplanet undergoes a dynamical collapse (phase 2) that results in a much hotter and compact planetary object (phase 3). Shown in figure \ref{distfig} are the planetary radius $R$ (a), central temperature $T_c$ (b), effective temperature $T_e$ (c), and luminosity $L$ (d). The colors correspond for different radial distances (see legend). As expected, the pre-collapse timescale increases as the radial distance decreases due to the higher disk temperature which prevents the protoplanet from losing its internal energy efficiently. At a radial distance of $\sim$10 AU it is found that the protoplanet cannot contract and evolve to become a gravitationally-bound planet, and the protoplanet is expected to dissipate in the disk.
This result is in agreement with previous studies that suggest that formation of giant planets by gravitational instability is unlikely to occur at small radial distances (e.g., Rafikov, 2007, 2009; Cai et al., 2010, Durisen et al., 2007). We therefore conclude that giant planets, if formed by gravitational instability, must form and remain at large radial distances during the first $\sim$ 10$^5-10^6$ years of their evolution. For our study the initial configurations of the protoplanets at small radial distances correspond to protoplanets that have been transported to these regions within the first $\sim 10^3-10^4$ years or less. If the protoplanets are significantly more compact at small radial distances they are not expected to be affected by the disk and are likely to evolve further (see Helled \& Bodenheimer, 2010 for evolution of isolated protoplanets). In this context it should be clear that our study concentrates on newly-formed protoplanets and not evolved and/or compact protoplanets. \par

At radial distances larger than about 10 AU the protoplanet can contract and evolve to undergo a dynamical collapse, but the evolution is clearly affected by the presence of the disk. 
The pre-collapse evolution timescale is found to range from about 3$\times 10^5$ years up to $10^6$ years for radial distances between 50 and 11 AU, respectively. Small radial distances result in an expansion of the planetary radius due to the higher disk's temperature. 
At radial distances of 10-20 AU the protoplanet is expanding, and once that sufficient energy is released, the protoplanet can contract further and reaches the dynamical collapse phase. The planetary expansion is found to be more significant for small radial distance, essentially up to the critical radial distance (10 AU in our case) in which the protoplanet is unable to contract and is expected to dissolve in the disk. Figure \ref{distfig}b shows the central temperature vs. time, it is shown that the central temperature increases on shorter timescales for larger radial distances. At 10 AU, where the protoplanet cannot contract, it can be seen from the figure that the central temperature actually decreases eventually, and the hydrogen dissociation point cannot be reached. As a result, we suggest that under these conditions the protoplanet cannot evolve to become a gravitationally bound object. 
The internal structures of the protoplanet throughout its evolution at 11 AU, the minimum radial distance allowing contraction, are  shown in figure \ref{11strctfig}.
\par

The planetary effective temperature and luminosity as a function of time are shown in figure \ref{distfig}c and \ref{distfig}d. It can be seen that at large radial distances the effective temperature increases slowly with time, while for smaller radial distances (smaller than $\sim$ 20 AU), the effective temperature remains constant, similar to the temperature of the surrounding disk. At large radial distances, due to the larger planetary radius and larger temperature difference  between the planetary outer layers and the disk, the initial luminosity is higher, and in addition, the temperature and pressure of the disk are significantly lower. As a result, the contraction occurs more rapidly, and the planetary luminosity decreases as the protoplanet evolves. Closer to the star, on the other hand, the protoplanet is slightly more compact and the surrounding temperature is significantly higher. Under these conditions the planetary contraction is less efficient and the protoplanet cannot release its energy effectively. This effect can be seen in figure \ref{distfig}d; the early stages of evolution at small radial distances are characterized by negative luminosity values (in these cases the effective temperature $T_e$ is actually the photospheric temperature), which results in an expansion. The larger planetary radius then leads to an increase in the luminosity which accelerates the energy release, and eventually leads to further contraction and dynamical collapse. \par

\begin{figure}[ftb]
\begin{center}
\centerline{\includegraphics[angle=0, width=10cm]{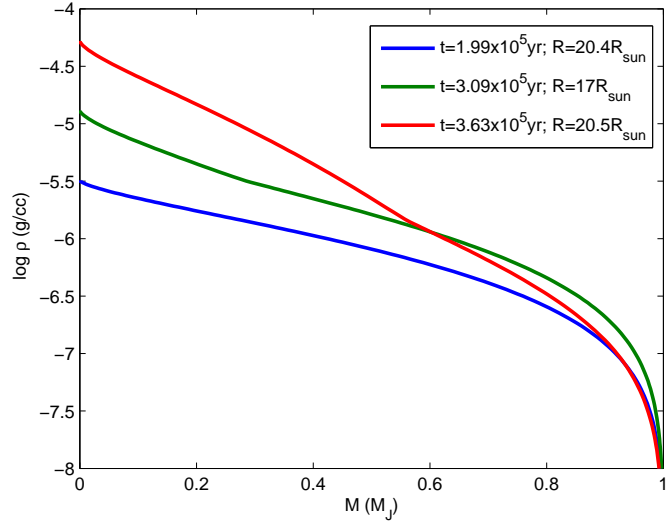}}
\caption{Density profile of 1 M$_J$ protoplanet at 50AU, for three different times near the collapse stage. The blue, green, and red curves correspond to $1.99\times10^5$, $3.09\times10^5$, and $3.63\times10^5$ years, respectively. From the difference between the green curve which presents a more compact configuration (R=17 R$_{Sun}$) and the red curve which corresponds to planetary expansion (R=20.5 R$_{Sun}$), it is clear that the inner structure becomes denser while the outer layers are thinner just before dynamical collapse occurs (see text for more details).}\label{expfig}
\end{center}
\end{figure}

\begin{figure}[ftb]
\begin{center}
\includegraphics[angle=0, width=10cm]{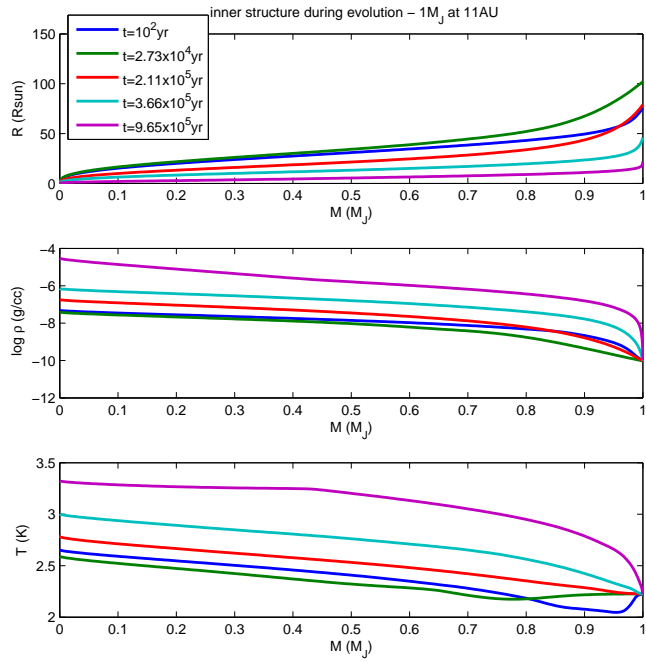}
\caption{Physical properties of a Jupiter-mass protoplanet at a radial distance of 11 AU at five different times throughout the pre-collapse evolution. Shown are the radius (R), density ($\rho$), and temperature (T) . At t=9.65$\times 10^5$ the protoplanet is on the edge of the dynamical collapse.}
\label{11strctfig}
\end{center}
\end{figure}

As shown in figure \ref{distfig}a, the planetary radius slightly increases just before the dynamical collapse occurs. This effect is caused by the dissociation of molecular hydrogen at the center, while the high internal temperatures lead to concentration of mass close to the center and an expansion of the envelope.  
The high central temperatures ($\sim$ 2000 K) result in an opacity decrease (due to grain evaporation), and the protoplanet changes from mostly convective to mostly radiative. 
This effect is demonstrated in figure ~\ref{expfig}, where the internal density structure at three different times near the dynamical collapse for 1M$_J$ protoplanet at 50AU are presented. The blue, green, and red curves correspond to times of $1.99\times10^5$, $3.09\times10^5$, and $3.63\times10^5$ years, respectively. 
At about 3$\times10^5$ years (green curve) the protoplanet contraction stops and the protoplanet then expands just before it reaches the dynamical collapse at  $t=3.63\times10^5$ years (red curve). It is clear from the figure, that at time $3.63\times10^5$ years the inner region becomes much denser and the outer layers expand. Just before the dynamical collapse is reached, there is a significant increase in the density near the center while the density closer to the surface decreases (due to the expansion in radius). \par 

\begin{figure}[ftb]
\begin{center}
\centerline{\includegraphics[angle=0,width=12cm]{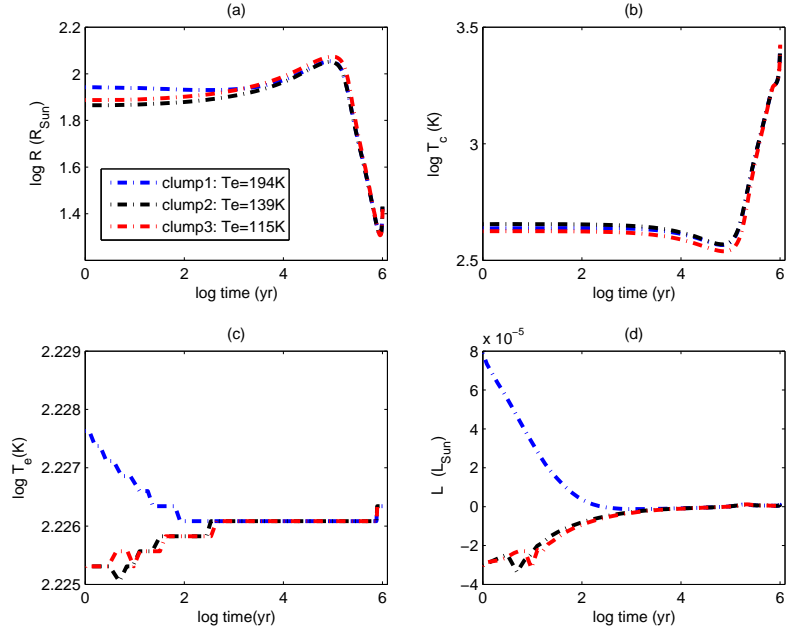}}
\caption{Pre-collapse evolution of a Jupiter-mass protoplanet at 11 AU for three different initial configurations. The blue, red, and black curves correspond to initial effective temperatures of about 195, 140 K and 115 K, respectively. Shown are the planetary radius $R$, central temperature $Tc$, effective temperature, and luminosity $L$. As can be seen from the figure, after $\sim$10$^2$ years the evolutionary tracks are essentially identical suggesting that the initial configuration has negligible effect on the global pre-collapse evolution. }\label{11evfig}
\end{center}
\end{figure}

\begin{figure}[ftb]
\begin{center}
\centerline{\includegraphics[angle=0,width=12cm]{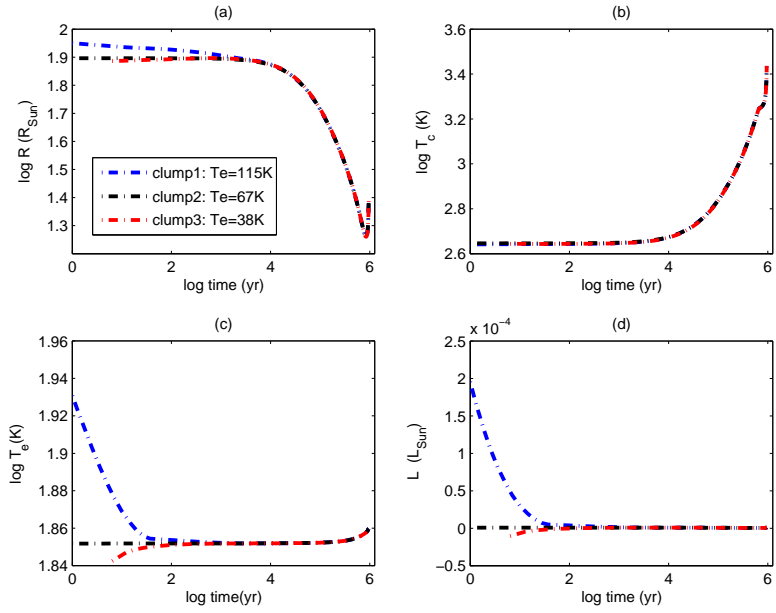}}
\caption{Same as figure \ref{11evfig}, but for a 1 M$_J$ protoplanet at 20 AU. 
The blue, red, and black curves correspond to initial effective temperatures of about 115, 70 and 40 K, respectively.}
\label{20evfig}
\end{center}
\end{figure}

In order to demonstrate the that the global pre-collapse evolution is not affected much by the initial conditions, we follow the evolution of a Jupiter-mass protoplanet at 11 AU, the minimum radial distance in which the protoplanet was found to evolve and contract, and also at $20$ AU, where the disk's  influence is weaker, starting with different initial configurations. The results for a Jupiter-mass at 11 AU are presented in figure \ref{11evfig}. The initial configuration for the protoplanet at 11 AU was originally chosen to have an effective temperature $Te$=139 K, and R$\sim$73 $R_J$ (black curves). We therefore consider also hotter (blue curves) and colder (red curves) initial configurations. 
As can be seen in the figure, after about 200 years there is no difference in the planetary evolution. 
For comparison, we also investigate the evolution at 20 AU using different initial effective temperatures and initial radii. The results are shown in figure \ref{20evfig}. In that case, due to the lower pressure and temperature of the disk the differences between the three evolutionary tracks are even smaller and are "forgotten" already after about 100 years. \par

We can therefore conclude that the planetary evolution is primarily affected by the planetary mass and the boundary conditions imposed by the disk (i.e., its temperature and pressure). The initial conditions are found to have only a small effect on the global pre-collapse evolution, and as a result, do not affect the resulting minimum radial distance found for clump's survival. 

\subsubsection{Evolution of Other Planetary Masses}\label{mss}

\begin{figure}[ftb]
\begin{center}
\centerline{\includegraphics[angle=0,width=10cm]{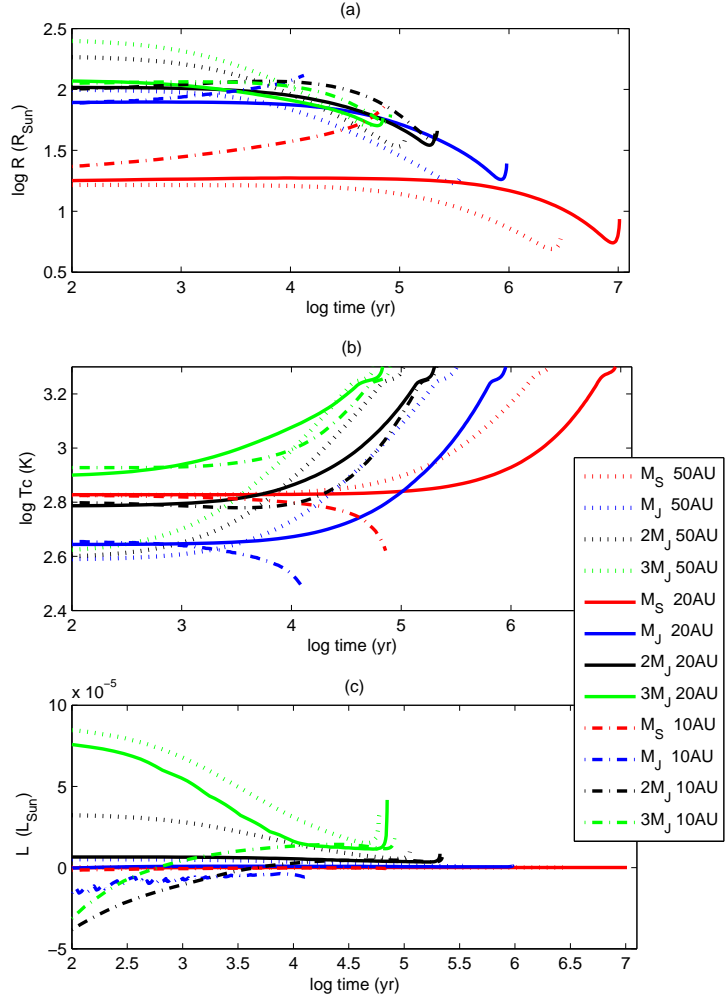}}
\caption{Pre-collapse evolution for Saturn-mass M$_S$ (red), 1 M$_J$ (blue), 2 M$_J$ (black), and 3 M$_J$ (green) at three radial distances: 10 AU (dashed curves), 20 AU (solid curves), and 50 AU (dotted curves). 
Shown are the planetary radius $R$, central temperature $Tc$, and luminosity $L$.}\label{massfig}
\end{center}
\end{figure}

The minimum distance of $\sim$ 11 AU for clump survival that we find above corresponds to 1 M$_J$ protoplanet. Clearly, more (less) massive protoplanets, due to larger (smaller) gravitational field, are expected to survive at smaller (larger) radial distances. 
Below, we repeat the calculation for protoplanets with masses of 1 M$_S$ (M$_S$ is Saturn's mass), 2 and 3 M$_J$. The initial configurations for the various masses are given in table 1. The evolution of  the protoplanets at radial distances of 10, 20, and 50 AU is presented in figure  ~\ref{massfig}. As expected, the pre-collapse timescale is shorter for more massive protoplanets, and protoplanets with masses larger than 1 M$_J$ can still evolve at radial distance of 10 AU. It is clear from the figure that more massive planets have shorter pre-collapse timescales, and as a result, more massive protoplanets could reach dynamical collapse, and evolve to become bound objects at smaller radial distances.

A Saturn-mass protoplanet is found to dissolve at radial distance of $\sim$ 12 AU. 3 M$_J$ protoplanet dissolves at about 9 AU, and 5 M$_J$ (not presented) at $\sim$ 7 AU (with disk's temperature $>$ 300 K). In addition, it is found that more massive protoplanets are less influenced by the disk, and therefore, the differences in the pre-collapse evolution decrease with increasing planetary mass.
\par

\subsection{Other Disk Models}

The disk model we use based on the work of Bell et al. (1997) with $\dot{M}=10^{-6} M_{Sun}/yr$ and $\alpha=10^{-2}$ ({\it disk1}) is found to fit well with observations of young protoplanetary disks (e.g., Hartmann et al. 2005). However, since the properties of the disk influence the evolution of the protoplanets we consider other disk models and investigate the effect on the planetary evolution. \par

Given that gravitational instabilities are likely to occur in cold and/or dense disks 
we consider two more disk models that might be more appropriate in the context of giant planet formation in the disk instability scenario. For the second disk model, {\it disk2}, we set an accretion rate $\dot{M}=10^{-7} M_{Sun}/yr$ and $\alpha=10^{-2}$, which corresponds to a colder disk. The third disk model, {\it disk3}, represents a denser disk and is taken to be a factor of five denser than {\it disk1} but with a similar temperature profile. The physical properties of {\it disk2} and {\it disk3} in comparison with {\it disk1} at 50 and 20 AU  are listed in table 2. Also shown are the initial radii, effective and central  temperatures, and central densities of the protoplanets at the corresponding radial distances. \par

\begin{figure}[ftb]
\begin{center}
\centerline{\includegraphics[angle=0, width=10cm]{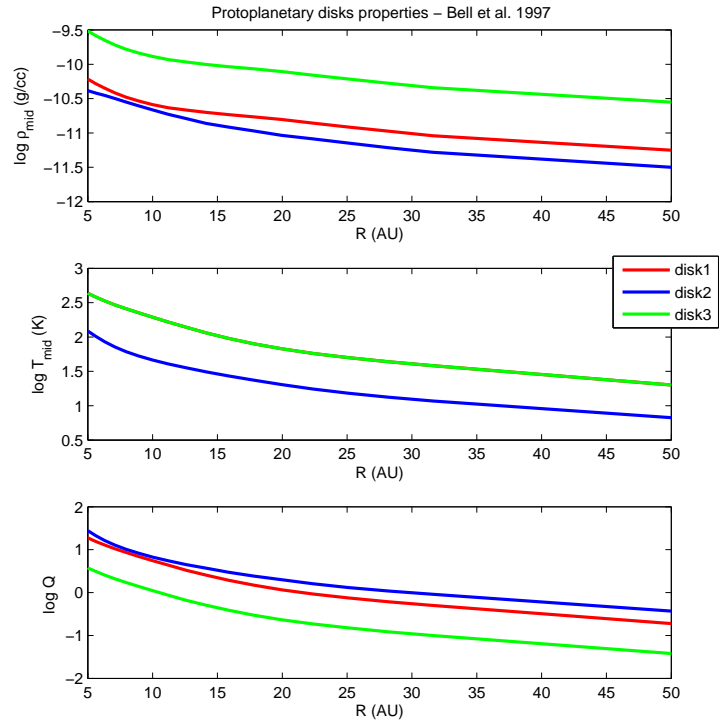}}
\caption{Physical properties of the three disk models we consider. Shown are the density, temperature and Toomre Q. The red, blue, and green curves correspond to {\it disk1, disk2} and {\it disk3}, respectively.} \label{disksfig}
\end{center}
\end{figure}

The physical properties of the three disks are shown in figure \ref{disksfig}. Shown are the midplane temperature, density, and Toomre Q between radial distances of 5 and 50 AU. Since the evolution model is 1D, there is only one representative disk temperature (and density) at a given radial distance.  It should be noted that the disk models are not gravitationally unstable in all the radial locations we consider. At fairly small radial distances the local Toomre $Q$ can be fairly high, therefore, in the context of this work, the protoplanets which are located at small radial distances ($\lesssim$ 20 AU) could exist at these locations {\it only} if migration took place, and do {\it not} correspond to in situ formation. \par 

It should also be noted that the disk models used in this work do not account for the influence of the forming planets on the disk. 
A fragmented disk is expected to develop spiral arms in which the protoplanets are formed, and as a result, leads to a change in the physical properties of the disk (i.e., temperature, density).  However, given the complex nature of the problem (which includes different timscales, resolutions, and dimensions) using a static disk is a good start. 
\par

\begin{figure}[ftb]
\centerline{\includegraphics[angle=0, width=13cm]{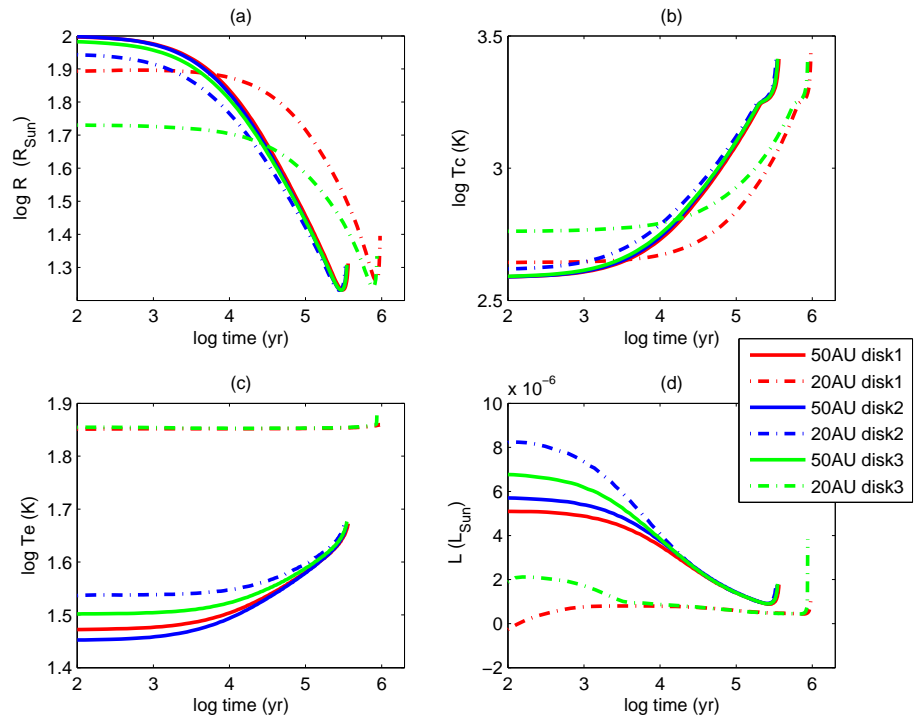}}
\caption{Pre-collapse evolution of a Jupiter-mass protoplanet for the three different disk models. The red, blue, and green curves correspond to {\it disk1, disk2} and {\it disk3}, respectively.}\label{diskevfig}
\end{figure}

The evolutions of a Jupiter-mass protoplanet at 20 and 50 AU for the three disk models are shown in figure \ref{diskevfig}. First of all, it is clear that the differences in the disk models are more dominant at 20 AU, and that the planetary evolution at 50 AU is essentially the same. For {\it disk2} due to the lower disk temperatures at 20 AU the protoplanet can contract more efficiently than in the case of {\it disk1}, resulting in a  pre-collapse evolution that is shorter by $\sim$ 6$\times 10^5$ years. The initial model of the protoplanet for {\it disk3} is more compact due to the higher disk density (and pressure) but the fairly high disk temperatures (similar to those of {\it disk1}) result in a pre-collapse evolution of the order of $10^6$ years. \par

The differences in the pre-collapse evolution and planetary physical properties become more significant at smaller radial distances. Regarding the minimum radial distance for survival, the low temperatures of {\it disk2} allow a 1 M$_{J}$ to contract and evolve even at 5 AU, in comparison with {\it disk1} which prevents the protoplanet to evolve at a radial distance of about 10 AU. Using {\it disk3}, we find that a Jupiter-mass protoplanet can contract at minimal radial distance of 9 AU. However, it should be kept in mind, that the formation of protoplanets at small radial distances is not very likely (e.g., D'Angelo et al., 2010 and references therein), and that if the protoplanets form at the outer disk, by the time they reach small radial distances due to migration they are likely to be more compact than the protoplanets considered here. We suggest that since the presence of the disk plays an important role in the pre-collapse evolution, it is desirable that in future work disk and planetary evolution models will be computed self-consistently.
 \par

\section{Gas Accretion}\label{accrtn}

Protoplanets can accrete gas from the surrounding disk as they evolve. In this section, we investigate the effect of gas accretion on the planetary evolution of a Jupiter-mass protoplanet. The composition of the disk is assumed to be  similar to that of the protoplanet, and the gas is added at the protoplanet's outermost layer. We consider three gas accretion rates: 0.05, 0.005, and 0.0005 M$_{\oplus}$/yr, and investigate the evolution of the protoplanet at 20 and 50 AU. \par

The process of gas accretion acts to increase the planetary total mass by inserting the accreted mass to the outer layer, and in addition, affects the energy budget by adding half of the potential energy of the gas, as an upper bound for the kinetic energy of the accreted gas (e.g., Frank 2002). We find that different fractions of the potential energy lead to similar long-term evolution, and that the effect of the accretion energy lasts for only about 1000 years.  
An accretion rate of 0.05 M$_{\oplus}$/yr is fairly high, and by using this accretion rate over the planetary evolution timescale of 1 M$_J$ we find that the accreted mass exceeds the available mass within the Hill sphere. However, since the planetary feeding zone is estimated to be $\sim$ 4 times the Hill sphere radius (e.g., Pollack et al. 1996, Greenzweig and Lissauer 1990) accretion of large mass of gas is feasible. Clearly, if the protoplanet migrates and reaches locations which are still gas-rich, the maximum mass that can be accreted will be determined by the accretion rate, and not by the available mass in the feeding zone. A detailed investigation of gas accretion with planetary migration is desirable to put tighter constraints on the gas accretion rate and we hope to address this topic in future research. \par   

\begin{figure}[ftb]
\centerline{\includegraphics[angle=0, width=13cm]{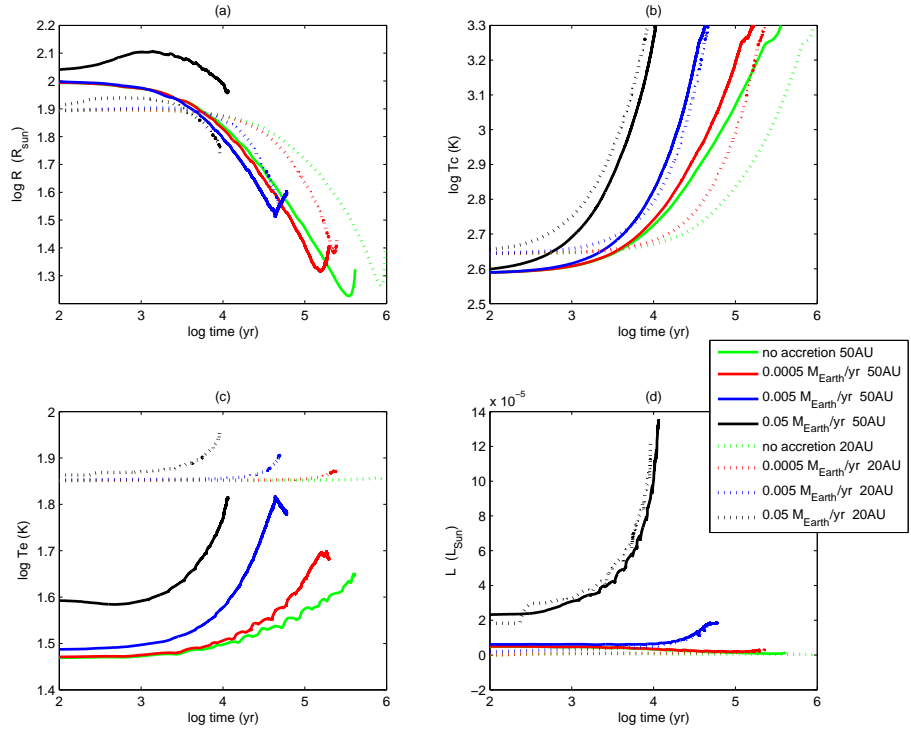}}
\caption{Pre-collapse evolution of a Jupiter-mass protoplanet with gas accretion rates of 0.0005M$_{\oplus}$/yr, (red) 0.005M$_{\oplus}$/yr (blue) and 0.05M$_{\oplus}$/yr (black) at 20 AU (dotted curves) and 50 AU (solid curves). 
For comparison, the planetary evolution at these two radial distances without gas accretion is also shown (green curves). } \label{acc1fig}
\end{figure}

The planetary evolution for the various accretion rates is shown in figure ~\ref{acc1fig}, the planetary radius $R$, central temperature $T_c$, effective temperature $T_e$, and luminosity $L$, are presented. For comparison, the evolution when no gas is accreted is presented as well. It can be seen from the figure that an increase of mass due to efficient gas accretion results in a faster contraction, as more massive protoplanets have shorter collapse timescales (Bodenheimer et al., 1980, Helled and Bodenheimer, 2010).\par

In addition, it is found that the planetary location has no significant influence on the planetary evolution when accretion is included. Gas accretion is found to change the planetary evolution by up to 20\%.  
For example, an accretion rate of 0.05 M$_{\oplus}$/yr (black curves) results in a difference of $\sim$ 2000 years in the total pre-collaspe timescale at radial distances of 20AU and 50AU. An accretion rate of  0.0005 M$_{\oplus}$/yr results in a change of about $5 \times 10^5$ years for these two radial distances.  
High accretion rates lead to significantly higher luminosity, due to the addition of the accretion energy, and the increase of mass. For these cases it is found that the protoplanet contracts on shorter timescales. \par

As discussed earlier, the increasing mass of the accreting protoplanet shortens the precollapse phase, as more massive planets contract on shorter timescales. However, the more rapid evolution when gas is accreted may not be only due to the addition of mass. We therefore compare the evolution of initial mass of 1 M$_J$ with gas accretion and a non-accreting protoplanet with an initial mass equals to the final mass when accretion is included. This way we can compare two formation scenarios which result in planets with similar final masses but with different pre-collapse evolutions. 

\begin{figure}[ftb]
\centerline{\includegraphics[angle=0, width=13cm]{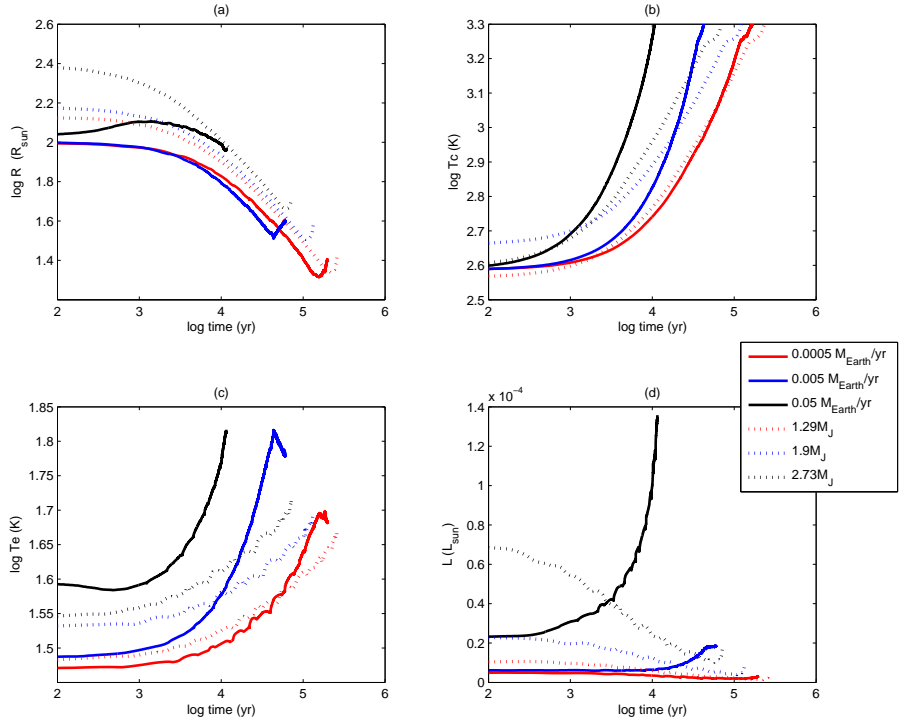}}
\caption{Pre-collapse evolution of a Jupiter-mass protoplanet with gas accretion rates (solid curves) of 0.0005M$_{\oplus}$/yr, (red) 0.005M$_{\oplus}$/yr (blue) and 0.05M$_{\oplus}$/yr (black). 
The final masses for accretion rates of 0.0005, 0.005 and  0.05 M$_{\oplus}$/yr  are 1.29 M$_J$, 1.9 M$_J$, and 2.73 M$_J$, respectively. Also presented is the evolution without accretion (dotted curves) of protoplanets with initial masses of 1.29 M$_J$ (red), 1.9 M$_J$ (blue), and 2.73 M$_J$ (black).
All evolutionary tracks are for radial distance of 50AU. } \label{acc2fig}
\end{figure}

The differences in the pre-collapse evolution for the two cases are shown in figure \ref{acc2fig}. 
The dotted curves correspond to non-accreting protoplanets with initial masses of 1.29 M$_J$, 1.9 M$_J$, and 2.73 M$_J$. These masses are the final masses derived for accretion rates of 0.0005, 0.005, and 0.05 M$_{\oplus}$/yr, respectively. The initial configurations of these planets are taken to be adiabatic, with the effective temperature similar to that of the disk. Under these assumptions, the initial radii of the protoplanets can be significantly larger. Gas accretion is found to shorten the precollapse phase due to the increase of mass together with the fairly compact configuration, which results in a denser and hotter protoplanet. Since gas is accreted as the protoplanet evolves, and at the same time, the initial radius of the protoplanet is small, the dynamical collapse is reached on short timescales. \par

It is interesting to note the different trends of the luminosity in the two cases (figure \ref{acc2fig}d). While for the accreting protoplanets the luminosity increases with time, the luminosity for the non-accreting protoplanets is decreasing. Clearly, the higher the accretion rate, the larger the difference in the pre-collapse timescale between accreting and non-accreting protoplanets.  \par

\section{Summary and Discussion}\label{concl}

We model the pre-collapse evolution of newly-formed protoplanets when the presence of the protoplanetary disk, and gas accretion are included. 
First, we concentrate on the change of the pre-collapse timescale of a Jupiter-mass protoplanet for various radial distances for a disk model with $\dot{M}=10^{-6} M_{Sun}/yr$ and $\alpha=10^{-2}$ ({\it disk1}), and find that a Jupiter-mass protoplanet cannot evolve and contract at radial distances smaller than about 11 AU due to the influence of the disk. 
While the exact radial distance for which contraction is no longer possible depends on the assumed disk model, it is clear from the models that the presence of the surrounding disk has an important effect on the planetary evolution, and that the pre-collapse timescale can be significantly longer at smaller radial distances. When the pressure and temperature of the disk are high enough, the protoplanet cannot contract to reach a dynamical collapse and become a gravitationally bound object. \par

We therefore suggest that contraction of protoplanets formed by gravitational instability is feasible only at relatively large radial distances in which the disk conditions do not affect the planetary evolution substantially. It is shown that the actual contraction timescale could vary substantially with the planetary location, and can change from 10$^5$ to $10^6$ years for 1 M$_J$, for radial distances between 50 and 11 AU, respectively. 
We then repeat the calculation for other planetary masses. A Saturn-mass protoplanet is found to dissipate at radial distance of $\sim$ 12 AU, while 3 M$_J$ and 5 M$_J$ protoplanets are found to  dissipate only at 9 AU, and 7 AU, respectively. It is also found that the influence/presence of the protoplanetary disk is less significant for more massive protoplanets. \par

We also investigate the sensitivity of the pre-collapse evolution for different disk models. It is found that the differences in the pre-collapse evolution are greater at smaller radial distances. While the density of the disk has a fairly modest effect on the planetary evolution ({\it disk3)}, the temperature of the disk has a significant effect on the planetary evolution,  and colder disks ({\it disk2}) are found to shorten the pre-collapse stage and allow planet evolution (and survival) at smaller radial distances. However, since giant planets in the disk instability scenario are likely to form at large radial distances, for which the difference in pre-collapse evolution is small, the pre-collapse evolution timescale for different disk models is not expected to change significantly. 
\par

Finally, we investigate the effect of gas accretion on the planetary evolution assuming three different accretion rates. It is shown that an increase of mass due to efficient gas accretion results in a faster contraction, as more massive protoplanets have shorter collapse timescales. It is also found that the planetary location has no significant influence on the planetary evolution when high accretion rates are considered. High accretion rates lead to significantly higher luminosities which result in shorter pre-collapse timescales. 
From a comparison between accreting and non-accreting protoplanets with the same final mass, we demonstrate the strong effect of gas accretion on the pre-collapse evolution; the higher the accretion rate, the greater the difference in the pre-collapse timescale. Accreting protoplanets reach the dynamical collapse stage significantly faster due to a more compact initial configuration and higher luminosity. The difference in timescale between accreting and non-accreting protoplanets with the same final mass can lead to significant differences in the final planetary structure and bulk composition due to different efficiency in planetesimal accretion and core formation (e.g., Helled et al., 2006, 2008). \par 

Our work demonstrates the importance of including the presence of the disk and gas accretion on the pre-collapse evolution of young protoplanets. This study suggests that there is no "standard" pre-collapse timescale and physical configuration for a given planetary mass as these can change significantly. As  a result, the pre-collapse timescale cannot be simply scaled (inverse) proportionally to the planetary mass. The pre-collapse stage has important implications for the planetary final composition (e.g., Helled and Bodenheimer, 2010, 2011; Boley et al., 2010), its survival, and interaction with the disk, and it is therefore important to further investigate how this evolutionary stage is affected by the planetary environment and other physical processes. 

The disk instability model for giant planet formation is more efficient at large radial distances, where the contraction timescale is found to be shorter. Even if giant planets can form close to the star, we suggest that due to the presence of the disk, the clumps are likely to dissipate. However, if the protoplanets are formed at large radial distances and migrate inward {\it after} the dynamical collapse, when the protoplanets are denser and more compact, the planets could survive at small radial distances. Alternatively, protoplanets could form and survive at small radial distances if their initial masses are sufficiently large (i.e., $>>$ 1 M$_J$). 
This conclusion is in good agreement with previous research, but for the first time, is based on planetary evolution considerations. Clearly, formation models that include planetary evolution and migration self-consistently are required to provide a more complete picture on planetary formation and evolution in the gravitational instability scenario. Finally, it will be interesting to include heavy elements in the models and investigate whether gaseous protoplanets can indeed become super-Earths/mini-Neptunes due to photo-evaporation of their gaseous envelopes and/or tidal stripping   (Boss et al., 2002; Nayakshin, 2011, Boley et al., 2010). The presence of heavy elements also affects the planetary opacity, and therefore the evolution, in particular, grain growth and settling have important impact on the timescale of the pre-collapse stage and the final intern structure (see Helled and Bodenheimer, 2011 for details). Clearly, other physical processes such as planetary migration, planetesimal capture, rotation, etc. could also affect the pre-collapse evolution, and we hope to include them in future research.  \par

\subsection*{Acknowledgments}
We thank A. Kovez for valuable and fruitful discussions and technical support. We also acknowledge helpful suggestions from M. Podolak, and P. Bodenheimer. 
This research was supported by the Israel Science Foundation (grant No. 1231/10).

\newpage
\section*{References}

Bell, K. R., Cassen, P. M., Klahr, H. H. and Henning, Th., 1997. ApJ, 486, 372\\
Boley, A. C., 2009. ApJ, 695, L53 \\
Boley, A. C.,  Hayfield, T., Mayer, L., Durisen, R. H., 2010. Icarus,  207, 509 \\
Boley, Aaron C.; Helled, Ravit; Payne, Matthew J., 2011. ApJ,  735, 30 \\
Boss, A. P., 1997. Science 276, 1836 \\
Boss, A. P., 2002. ApJ. 576, 462 \\
Boss, A. P., Wetherill, G. W. and Haghighipour, N., 2002. Icarus, 156, 291 \\
Boss, A. P., 2011. ApJ, 731, 74 \\ 
Cai, K., Durisen, R. H., Boley, A. C., Pickett, M. K., and Mej??a, A. C. 2008. ApJ, 673, 1138 \\
Cai, K., Pickett, M. K., Durisen, R. H., Milne, A. M., 2010. ApJ, 716, L176\\
D'Alessio, P., Calvet, N., Hartmann, L., Lizano, S., Cant{\'o}, J., 1999. ApJ, 527, 893\\
Decampli, W. M., Cameron, A. G. W., 1979. Icarus 38, 367\\
D'Angelo, G., Durisen, R. H., Lissauer, J. J., 2011. Giant Planet Formation.  Exoplanets, edited by S. Seager. Tucson, AZ: University of Arizona Press\\ 
Durisen, R. H., Boss, A. P., Mayer, L., Nelson, A. F., Quinn, T., and Rice, W. K. M. 2007.  in Protostars and Planets V, 607 \\
Durisen, R. H., Hartquist, T. W., Pickett, M. K., 2008. Astrophysics and Space Science, 317, 3 \\
Frank, J., King, A., Raine, D., 2002. Accretion power in astrophysics, third addition. Cambridge university press \\
Greenzweig, Y., Lissauer, J.J., 1990. Icarus, 87, 40 \\
Hartmann, L., Megeath, S.~T., Allen, L., et al., 2005. ApJ, 629, 881\\
Hayfield, T., Mayer, L., Wadsley, J., and Boley, A. C., 2011.  MNRAS, 417, 1839\\
Helled, R., Podolak, M., Kovetz, A., 2006. Icarus 185, 64\\
Helled, R., Podolak, M., Kovetz, A., 2008. Icarus 195, 863\\
Helled, R. and Bodenheimer, P., 2011. Icarus, 207, 503\\
Helled, R. and Bodenheimer, P., 2011. Icarus, 211, 939\\
Hubickyj, O., Bodenheimer, P., Lissauer, J. J., 2005. Icarus 179, 415\\
Kovetz, A., Prialnik, D., and Shara, M. M., 1998. ApJ, 325, 828\\
Kovetz, A., Yaron, O., and Prialnik, D., 2009. MNRAS, 395:1857-1874\\
Mayer, L., Lufkin, G., Quinn, T.,   Wadsley, J., 2007. ApJ, 661, L77\\
Mayer, L., Quinn, T., Wadsley, J.,  Stadel, J., 2004. ApJ, 609, 1045\\
Mayer, L., Quinn, T., Wadsley, J.,  Stadel, J., 2002. Science, 298, 1756\\
Nayakshin, S., 2010. MNRAS, 408, L36\\ 
Nayakshin, S., 2011. MNRAS, 416, 2974\\
Pollack, J., McKay, C., Christofferson, B., 1985. Icarus 64, 471\\
Pollack, J. B., Hubickyj, O., Bodenheimer, P., Lissauer, J. J., Podolak, M., and Greenzweig, Y., 1996. Icarus 124, 62\\
Rafikov, R. R., 2007. ApJ, 662, 642\\
Rafikov, R. R., 2009. ApJ, 704, 281\\
Rice, W. K. M., Lodato, G., Pringle, J. E., Armitage, P. J., Bonnell, I. A., 2004. MNRAS, 355, 543\\

\clearpage
\begin{table}
\center
\label{inittab}
\begin{tabular}{l c c  c c |} 
 \hline  \hline 
Disk Properties: $a=50$ AU &T = 20 K &logP = -2.56 \\  
\hline
Mass (M$_J$)& Radius (R$_{Sun}$)& T$_e$ (K)& T$_c$ (K) \\
\hline
0.2987&16.4&27&671\\
1&99.6&29&387 \\
2&194.6&32&447 \\
3&273.7&35&502 \\
5&355.7&42&637 \\
 \hline  \hline 
Disk Properties: $a$ = 20 AU &T = 67.5 K &logP = -0.84 \\  
\hline
Mass (M$_J$)& Radius (R$_{Sun}$)& T$_e$ (K)& T$_c$ (K) \\
\hline
0.2987&18.4&70&672\\
1&78.7&71&441 \\
2&102.1&71&623 \\
3&108.2&74&869 \\
5&116&82&1349 \\
\hline
\hline
\end{tabular}\\
\caption{Initial clump properties at two radial distances (50 and 20 AU). The disk model ({\it disk1}) is taken from Bell et al., (1997). 
}
\end{table}

\clearpage 
\begin{table}
\center
\label{disktab}
\begin{tabular}{c  c  c  c | c  c  c  c} 
 \hline
 \hline
 \hline
 \multicolumn{4}{c}{Disk Properties}& \multicolumn{4}{c}{Protoplanet Initial Model - 1 M$_J$}\\  
  &  a (AU) & T (K) & logP (cgs) & Radius (R$_{Sun}$) & T$_e$ (K) & T$_c$ (K) & $\rho_c$ (g/cc) \\
\hline
 {\it disk1} & & & & & & & \\
 \hline
 & 50& 20& -2.56& 99.6& 29& 405& $3.5\times 10^{-8}$ \\
 & 20& 67.5& -0.84& 78.7& 71& 443& $4.6\times 10^{-8}$ \\
\hline {\it disk2} & & & & & & & \\
 \hline
 & 50& 6.7& -3.29& 100.2& 28& 392& $3.2\times 10^{-8}$ \\
 & 20& 20.3& -1.59& 88.4& 34& 442& $4.6\times 10^{-8}$ \\
\hline {\it disk3} & & & & & & & \\
 \hline
 & 50& 20& -1.86& 96.5& 32& 411& $3.7\times 10^{-8}$ \\
 & 20& 67.5& -0.14& 53.7& 71& 584& $1.0\times 10^{-7}$ \\
\hline
\hline
\hline
\end{tabular}\\
\caption{Initial clump properties at 50 and 20 AU for the three disk models we consider ({\it disk1, disk2, disk3}). See text for details.}
\end{table}

\clearpage 
%

\clearpage

\clearpage

\end{document}